\documentclass[aps,pre,twocolumn,showpacs,groupedaddress]{revtex4}

\usepackage{graphicx}

 \newcommand {\dip}{\displaystyle}
 \newcommand {\s}{\dip \sqrt}
 \newcommand {\f}{\dip \frac}
 
 \newcommand {\ig}{\includegraphics}

\begin{document}


\title{Escape time statistics for mushroom billiards}


\author{Tomoshige Miyaguchi}
 \email{tomo@nse.es.hokudai.ac.jp}
\affiliation{%
Meme Media Laboratory, 
Hokkaido University, 
Kita-Ku, Sapporo 060-0813, Japan 
}%


\date{\today}

\begin{abstract}
 Chaotic orbits of mushroom billiards display intermittent
 behaviors. We investigate statistical properties of this system by 
 constructing an infinite partition on the chaotic part of a Poincar\'e 
 surface which illustrates details of chaotic dynamics. Each 
 piece of the infinite partition has an unique escape time from the half
 disk region, and from this result it is shown that, for fixed values
 of the system parameters, the escape time distribution obeys  power law  
 $1/t_{\rm  esc}^3$.   
\end{abstract}


\maketitle


\section{Introduction}

Fully chaotic dynamical systems such as the baker transformation and the
Arnold's cat map are statistically characterized by, for example,
exponential decay of correlation functions with decay rates given by
the Pollicotte--Ruelle resonances (See Ref.~\cite{ruelle1} and references
therein) and exponentially fast escape from regions of phase
spaces with the escape rate given by the positive Lyapunov exponents and
KS (Kolmogolov--Sinai) entropy \cite{gaspard,dorfman}. These properties
are outcomes of the  uniform hyperbolicity, which means the uniform
instability of chaotic trajectories.  

In contrast to such ideally chaotic systems,  phase spaces of
generic Hamiltonian systems consist not only of non-integrable chaotic
regions but also of integrable regions (torus), where motions are
quasi-periodic \cite{lichtenberg}, and therefore the uniform instability
may be broken in these systems. In fact, generic Hamiltonian systems
frequently exhibit power law type behaviors, which is due to occasional
trappings of chaotic orbits in neighborhoods of torus. Although these
phenomena are observed in many systems 
\cite{karney,mackay,chirikov,geisel,solomon}, analytical derivations of
decaying properties of correlation functions and sticking time
distributions are difficult mainly  because there exist complex fractal 
torus structures.  

In order to understand power law behaviors in dynamical systems,
non-hyperbolic 1-dimensional mappings have been studied by several
authors (e.g.,~Refs. \cite{pomeau1,pomeau2,geisel1,aizawa,tasaki1}).  
Therefore it is natural to imagine connections of these non-hyperbolic
maps and mixed type Hamiltonian systems, however, extensions of these
maps to 2-dimensional area-preserving systems are unknown (but see
Refs.~\cite{artuso2,tm}). Thus it is important to elucidate the
properties of non-hyperbolicity which is typical in the mixed type
Hamiltonian systems.    

The mushroom billiard, which has been proposed by Bunimovich recently
\cite{bunimovich}, is expected to be a candidate of analytically
tractable model for such problems of mixed type systems. This is because
the mushroom billiard system does not have the fractal torus structures 
and chaotic and torus regions are sharply divided. (See
Ref.~\cite{prosen}, for other example of such systems.) 
Thus the mushroom billiard system can be thought as an ideal model for
understanding mixed type Hamiltonian systems, and it has already been 
under active researches \cite{bunimovich2,altmann,shudo,dietz}. 

In this paper, we give a theoretical derivation of the escape time
distribution for fixed values of the system parameters. In
Ref.~\cite{altmann}, it has already been shown numerically that it obeys  
power law, and our result agrees theirs perfectly.  In order to derive
the escape time distribution, we begin with the construction of an
infinite partition on a Poincar\'e surface, which reveals detailed
dynamics in neighborhoods of the outermost tori.

This paper is organized as follows. In Sec.~II, we introduce the
mushroom billiard system, and define a Poincar\'e map and its inverse 
transformation. In Sec.~III, we construct the infinite partition 
by using the inverse of the Poincar\'e map recursively. And in Sec.~IV,
the escape time distribution is derived from the structure of the
infinite partition.  A brief discussion is given in Sec.~V.

\begin{figure}
\vspace*{.2cm}
\ig[width=5.7cm,height=5.7cm]{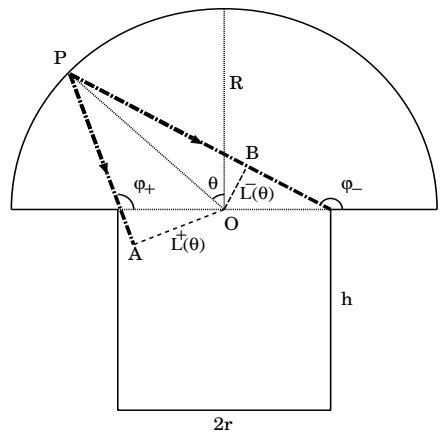}
 \caption{\label{fig:1st-exit-board} The shape of the board of the
 mushroom billiard (the solid lines), which consists of a half disk (the
 hat) and a rectangle (the foot). A point particle inside the board
 moves freely except for the elastic collisions with the walls. The
 absolute value of the angular momentum $|L|$ equals to the distance
 between the origin and the trajectory (See Sec.~II A.).  The boundaries
 $L^{\pm}(\theta)$ of the first escape domain 
 $L^-(\theta) < L < L^+(\theta)$ are also displayed (See Sec.~II B.). }     
\end{figure}

\section{Poincar\'e Map and Its Inverse}

The mushroom billiard is defined by the motion of a point particle on
the billiard board depicted in Fig.~\ref{fig:1st-exit-board}. This board   
consists of a half disk (the hat) of radius $R$ and a rectangle (the
foot) of width $r$ and height $h$~\cite{bunimovich}. 
We use the polar coordinate $(u, \theta)$, and the Cartesian coordinate
$(x,y)$; we set the origin as the center of the half disk in both
cases. The  angle variable $\theta$ is defined as the angle between the
position vector of the point particle and the vertical line (See
Fig.~\ref{fig:1st-exit-board}).   

\subsection{The definition of the Poincar\'e Map}

We define a Poincar\'e surface at the arc of the semicircle 
$x^2+y^2=R^2 (y\geq 0)$ with negative momentum of the radial 
direction, namely, just after the collision with the arc. For the 
coordinate of the Poincar\'e map, we use the angle $\theta$ and the 
associated angular momentum $L$. This Poincar\'e map $\Phi(L, \theta)$ 
is area-preserving; it can be proved through a direct  
calculation of the Jacobian of the map which equals to 1 
everywhere. This coordinate system is slightly different from the
Birkhoff coordinate, because the former is defined only on the arc, but
the latter on the whole boundary of the billiard board. 

We also set the kinetic energy as $v_x^2+v_y^2=1$. Although this setting
is not essential, it is convenient to calculate the angular momentum
$L$; the absolute value of the angular momentum $|L|$ equals to the
distance from the origin to the trajectory.
In Fig.~\ref{fig:1st-exit-board}, for example, if a point particle 
moves on the line $PA$ in the direction described in the figure, its
angular momentum $L$ ($L>0$) equals the length of the segment $AO$ (the
dashed line), and if a point particle moves on the line $PB$ in the
direction described in the figure, the absolute value of the angular
momentum $-L$ ($L<0$) equals the length of the segment $BO$ (the
long-dashed line). 

\begin{figure}
\hspace*{-0.cm}
\ig[width=7.5cm,height=5.5cm]{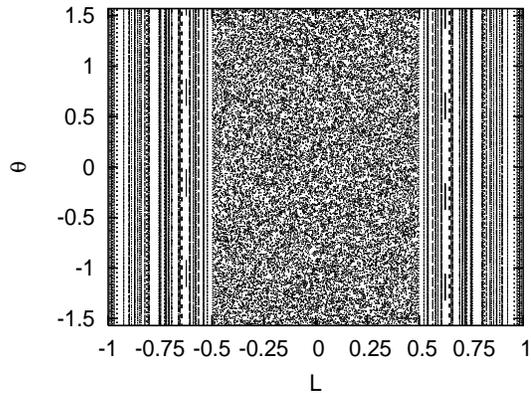}
 \vspace*{-.cm}
 \caption{\label{fig:ps-mush1} The Poincar\'e surface for $R=1$, 
 $~r=0.5$ and $h=1$.  The region $|L|<0.5$ is chaotic and the other
 integrable.} 
\end{figure}

We display an example of the Poincar\'e surface in
Fig.~\ref{fig:ps-mush1}. The Poincar\'e map $\Phi(L, \theta)$ is
defined on 
${\mathcal D} = \{(L,\theta) \in [-R,R]\times[-\pi/2,\pi/2]\}$;
the region $|L| < r$ is chaotic, and $|L| > r$ is torus. 
The Poincar\'e map is symmetric with respect to the origin 
$(L, \theta) = (0,0)$, i.e., 
$\Phi (L, \theta) = -\Phi(-L, -\theta)$.  In the subsequent subsections, 
we will restrict the domain of the Poincar\'e map to the region of the
negative momentum in order to simplify the analysis. 

\subsection{The first escape and re-injection domains}

Let us consider a point $(L, \theta)$ on the Poincar\'e surface such that the 
original orbit of the billiard system (the continuous time flow)
starting from this point escapes from the hat region to the foot
without no collision. We define the {\it first escape domain }$\mathcal D_1$
as all such points on the Poincar\'e surface. The boundary 
of $\mathcal D_1$ can be calculated analytically as follows; fix the
angle $\theta$ on the Poincar\'e surface, if the angular momentum $L$
satisfies the relation $L^-(\theta) < L < L^+ (\theta)$, then 
$(L, \theta) \in {\mathcal D_1}$, where $L^{\pm}(\theta)$ are displayed
in Fig.~\ref{fig:1st-exit-board}. More precisely, $L^{\pm}(\theta)$ are
defined as  
\begin{eqnarray}
\begin{array}{lll}
L^{\pm}(\theta) &=& \pm r \sin (\pi - \varphi_{\pm})  \\[.35cm]
&=& \pm r R \cos \theta ~(R^2 + r^2 \mp 2Rr \sin \theta)^{-1/2},
\end{array}
\label{eqn:1st-exit-L}
\end{eqnarray}
where the angles $\varphi_{\pm}$ are defined as in
Fig.~\ref{fig:1st-exit-board} and given by 
$\tan \varphi_{\pm} = R\cos \theta / (-R \sin \theta \pm r)$.
Next, let us consider the the domain with negative angular momentum,
$L^-(\theta) < L < 0$; the positive domain $L > 0$ can be treated in the
same way because of the symmetry. 
Solving the equation $(L^-(\theta))^2 > L^2$ in terms of $\theta$, the
first escape domain for $L<0$ can be represented as
\begin{eqnarray}
{\mathcal D_1} 
= \{(L, \theta) \in {\mathcal D^-}~|~\theta^-(L) < \theta < \theta^+ (L)\},
\label{eqn:1st-exit-theta1}
\end{eqnarray}
where ${\mathcal D^-} = \{(L,\theta) \in [-r,0]\times[-\pi/2,\pi,2]\}$
is the chaotic region with negative angular momentum and
$\theta^{\pm}(L)$ are defined as 
\begin{eqnarray}
\theta^{\pm}(L) \hspace*{-.05cm} = \hspace*{-.05cm}\arcsin \hspace*{-.05cm}
\left\{ \hspace*{-.05cm}
\f {-L^2 \pm \s {L^4 -L^2(R^2 + r^2) + r^2 R^2}}{rR}
\right\} \hspace*{-.1cm} .
\label{eqn:1st-exit-theta2}
\end{eqnarray}
The functions $\theta^{\pm}(L)$ defines the boundary of the first escape
domain $\mathcal D_1$. 

\begin{figure}
 \vspace*{-.cm}\hspace*{0.5cm}
\ig[width=8.0cm,height=6.7cm]{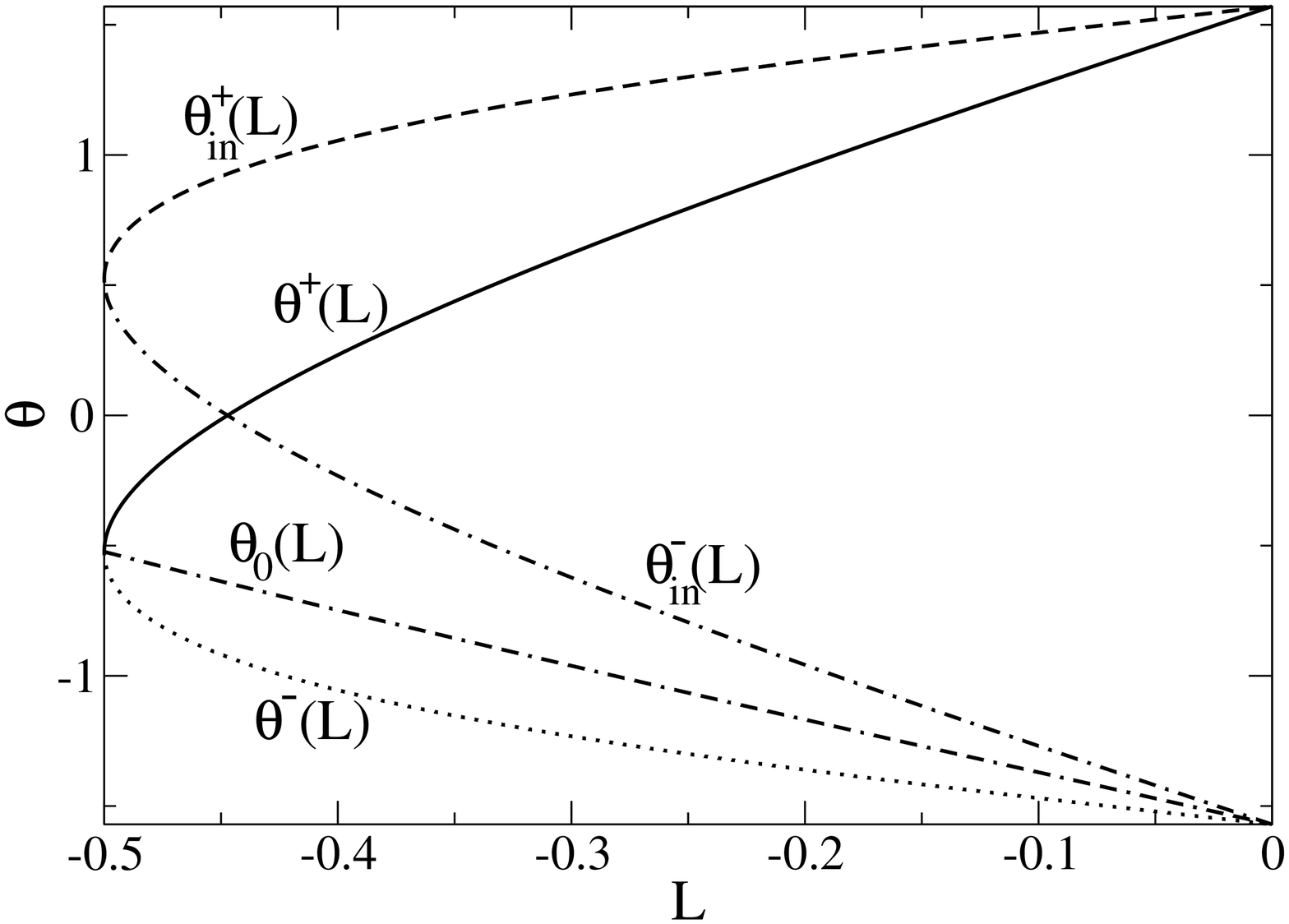}
 \vspace*{-.cm}
 \caption{\label{fig:1st-exit} The boundaries $\theta^+(L)$ (the solid
 line), $\theta^-(L)$ (the dotted line), $\theta^+_{\rm in}(L)$ (the
 dashed line), $\theta^-_{\rm in}(L)$ (the dotted-and-dashed line), and   
 $\theta_{0}(L)$ (the dotted-and-double-dashed line) are described. 
 We set the system parameters as $R=1$, $~r=0.5$ and $h>0$.
 The only negative part of the angular momentum $L<0$ is displayed
 because  the Poincar\'e surface is symmetric. }  
\end{figure}

An orbit of the billiard flows starting from the first escape domain
exits the hat region and stays in the foot for some times; and then it
returns to the hat and reaches again to the Poincar\'e surface. We
define the {\it re-injection domain} $\mathcal D_{\rm in}$ on the
Poincar\'e surface as all such just returning points , more precisely,
we define   
$
\mathcal D_{\rm in} := \Phi(\mathcal D_1)$. $\mathcal D_{\rm in}
$ 
can be derived in the same way as $\mathcal D_1$;
\begin{eqnarray}
{\mathcal D_{\rm in}} 
= \{(L, \theta) \in {\mathcal D^-} ~|~ 
\theta^-_{\rm in}(L) < \theta < \theta^+_{\rm in} (L)\},
\label{eqn:inj-dom1}
\end{eqnarray}
where the boundary $\theta^{\pm}_{\rm in}(L)$ is defined by 
\begin{eqnarray}
\theta^{\pm}_{\rm in}(L) = \arcsin 
\left\{
\f {L^2 \pm \s {L^4 -L^2(R^2 + r^2) + r^2 R^2}}{rR}
\right\}\hspace*{-.1cm} .
\label{inj-dom2}
\end{eqnarray}
In Fig.\ref{fig:1st-exit}, ~$\theta^{\pm}(L)$ and $\theta^{\pm}_{\rm
in}(L)$ are displayed for $R=1$, $~r=0.5$ and $h>0$.   

\subsection{The inverse of the Poincar\'e map}

\begin{figure}
 \vspace*{.75cm}\hspace*{-0.0cm}
\ig[width=8.55cm,height=4.5cm]{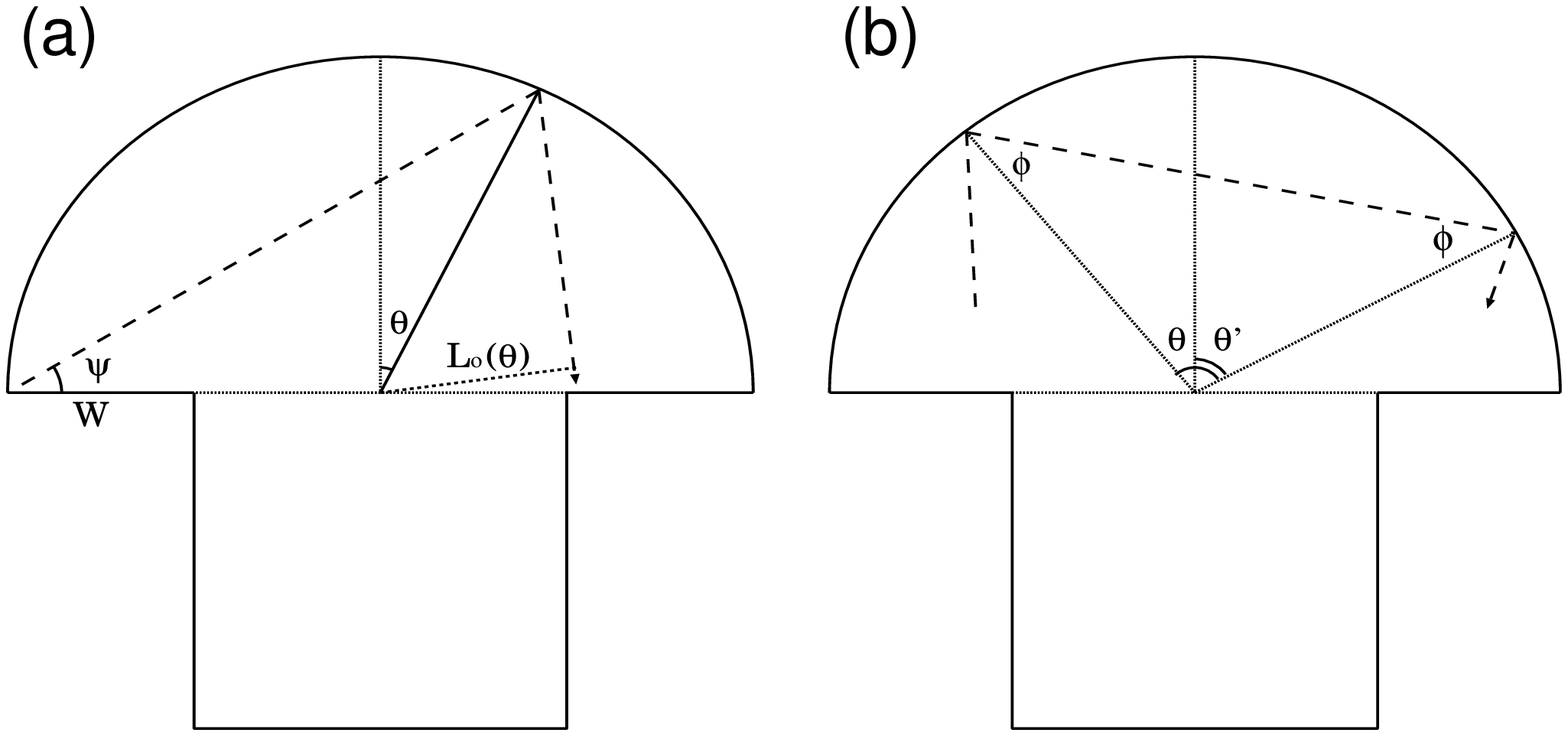}
 \vspace*{.3cm}
 \caption{\label{fig:mush-double} (a) A classification of orbits for
 fixed $\theta$ : if $L < L_0 (\theta) ~( < 0)$ there is a collision
 with the wall $W$ and if $L_0 (\theta) < L~( <  0)$ there is no
 collision. Note that we consider only $L<0$. In the  figure, we display
 the critical case $L=L_0(\theta)$ by the dashed  line. 
 (b) The inverse image of the Poincar\'e  map, which is derived from
 $\theta = \theta' + \pi - 2 \phi$ and $\phi = \arcsin|L/R|$.}
\end{figure}

When an orbit of the billiard flows collides with the boundary 
\begin{eqnarray}
W := \{~(x,y)~|~x \in [-R,-r]\cup [r, R] , ~y=0\},
\end{eqnarray}
the angular momentum changes its sign. We should take into account the 
collisions with this boundary $W$, because we reduce the Poincar\'e map
to the domain of negative momentum $L<0$. Let us consider the domain 
$\mathcal D^- \backslash {\mathcal D}_{\rm in}$ and its inverse image 
$\Phi^{-1}(\mathcal D^- \backslash {\mathcal D}_{\rm in})$. The original
orbit $l$ (namely flow) connecting a point   
$(L, \theta) \in \mathcal D^- \backslash {\mathcal D}_{\rm in}$ and 
$
\Phi^{-1} (L, \theta) 
\in \Phi^{-1}(\mathcal D^- \backslash {\mathcal D}_{\rm in})
$
are classified into two classes for fixed $\theta$ 
(See Fig.~\ref{fig:mush-double}(a).): when $L < L_0 (\theta) $, the
orbit $l$ is a line segment, namely there is no collision with the
boundary $W$ and when $L > L_0(\theta)$, the orbit $l$ consists of two
line segments, namely there is a collision with the boundary $W$. 
In Fig.~\ref{fig:mush-double}(a), an orbit for the critical case
$L=L_0(\theta)$ is displayed by the dashed line.
We can define $L_0(\theta)$ by 
\begin{eqnarray}
L_0(\theta) = -R \cos(\psi - \theta),
\end{eqnarray}
where $\psi$ is defined as depicted in
Fig.~\ref{fig:mush-double}(a). Furthermore, using  
$~\tan \psi = \cos \theta / (1 - \sin \theta)$, we have 
\begin{eqnarray}
L_0(\theta) = -R (2 -2 \sin \theta)^{-1/2} \cos \theta.
\end{eqnarray}
Solving the inequalities $L<L_0(\theta)$ and  $L>L_0(\theta)$ in terms
of $\theta$, we have the result:~when $\theta < \theta_0 (L) $ there is
no collision with the boundary $W$, and when $\theta > \theta_0(L)$
there is a collision with the boundary $W$, where $\theta_0(L)$ is
defined by 
\begin{eqnarray}
\theta_0 (L) = \arcsin \left( \f{2L^2}{R^2} - 1 \right).
\end{eqnarray}

Using these results and definitions, we can construct the inverse of the
Poincar\'e map $\Phi^{-1}$ on $\mathcal D^- \backslash D_{\rm in}$ as
follows (See Fig.~\ref{fig:mush-double}(b)), 
\begin{eqnarray}
\Phi^{-1}(L, \theta) = 
 \left\{ 
\begin{array}{ll}
\left(L, \theta + \pi - 2 \arcsin \left|  \f LR \right| \right), \\[.5cm]
&\hspace*{-.7cm}{\rm if}~ \theta < \theta_0 (L)
\\[.5cm]
\left(L, \theta       - 2 \arcsin \left| \f LR \right| \right), 
&\hspace*{-.7cm}{\rm if}~ \theta > \theta_0 (L)
\end{array}
\right.
\label{eqn:mush-inv}
\end{eqnarray}
Notice that the angular momentum is unchanged by the collisions with the
arc, and that we restrict the domain of the inverse map on the region
$L<0$ by identifying the points $(-L, -\theta)$ with $(L, \theta)$
when the point particle collides with the wall $W$, i.e., when 
$\theta > \theta_0 (L)$.

\section{The infinite partition}

Using this inverse map $\Phi^{-1}$, we can define the {\it $n$-th escape 
domain} $\mathcal D_n$ recursively:
\begin{eqnarray}
{\mathcal D_n} = \Phi^{-1} ({\mathcal D_{n-1} \backslash {\mathcal
 D_{\rm in}}} ), ~~~(n=2, 3, \cdots). 
\label{eqn:nth-dom}
\end{eqnarray}
Note that we should remove $\mathcal D_{\rm in}$ from $\mathcal D_{n-1}$
in the recursion relation Eq.~(\ref{eqn:nth-dom}), because the inverse
image of the injection domain $\Phi^{-1}({\mathcal D_{\rm in}})$ equals
to the first escape domain $\mathcal D_1$. 

We restrict the parameters as $R=1$, $r=0.5$ and $h>0$ in the
following and explicitly derive the boundaries of the $n$-th escape
domain $\mathcal D_n$. 
First, we derive explicitly the first to fourth escape domains, in order to
confirm that these four domains fill the domain $\mathcal D^-$ like
Fig.~\ref{fig:partition}(a) except for the three regions $E_1, E_2$ and
$ E_3$. Then the boundaries of the $n$-th escape domain ($n \geq 5$) can
be derived recursively.  

Let us start with $n=2$ of Eq.~(\ref{eqn:nth-dom}).
The domain ${\mathcal D_1 / \mathcal D_{\rm in}}$ can be divided into
three pieces:
\begin{eqnarray}
\begin{array}{lll}
{\mathcal D_1 / \mathcal D_{\rm in}}
&=& \{ (L, \theta)~|~\theta^-(L) < \theta < \theta_0(L) \} \\[.35cm]
&& \hspace*{-1.7cm} \cup
\{ 
(L, \theta)~|~\theta_0(L) < \theta < \theta^+(L) , ~-r < L < L^-(0)
\} 
\\[.35cm]
&& \hspace*{-1.7cm} \cup
\{ (L, \theta)~|~\theta_0(L) < \theta < \theta^-_{\rm in}(L) , 
~L^-(0) < L < 0 \}.
\end{array}
\label{eqn:division}
\end{eqnarray}
where $L^-(0)$ is defined by Eq.~(\ref{eqn:1st-exit-L}) (See also
Fig.~\ref{fig:partition}(b)). 
In Eq.~(\ref{eqn:division}), we abbreviate the expression 
$(L, \theta) \in {\mathcal D^-}$  to the one $(L, \theta) $ for
simplicity; we use the same abbreviation in what follows.
Let us represent the three sets on the right hand side as 
$\mathcal F^1_1$, 
$\mathcal F^1_2$, and $\mathcal F^1_3$, respectively; namely, 
$\mathcal D_1 / \mathcal D_{\rm in} = 
\mathcal F^1_1 \cup \mathcal F^1_2 \cup \mathcal F^1_3$.
These three sets are displayed in Fig.~\ref{fig:partition}(b). 
Using these notations and the inverse map $\Phi^{-1}$ 
[Eq.~(\ref{eqn:mush-inv})], the
second escape domain $\mathcal D_2$ is given 
by 
\begin{eqnarray}
\begin{array}{lll}
{\mathcal D_2}  &=&
\Phi^{-1} ({\mathcal D_{1} / {\mathcal D_{\rm in}}} ) \\[.35cm]
&=&
\Phi^{-1} (\mathcal F^1_1) \cup \Phi^{-1} (\mathcal F^1_2) \cup 
  \Phi^{-1} (\mathcal F^1_3).
\end{array}
\end{eqnarray}
Let us denote the $\theta$ component of $\Phi^{-1}(L, \theta)$ as 
$\Psi^L (\theta)$, and inverse image of $\mathcal F_1^1$ as 
$\mathcal F^2_1 \equiv \Phi^{-1} (\mathcal F^1_1) $. Using these
definitions, we have
\begin{eqnarray}
\begin{array}{lll}
\mathcal F^2_1 &=&
\{
(L, \theta)~|~\Psi^L(\theta^-(L)) < \theta < \Psi^L (\theta_0(L)) 
\} \\[.35cm]
&=& \{
(L, \theta)~|~\theta^-(L) + \pi - 2 \arcsin \left| L \right| 
< \theta < \frac {\pi}2)
\},
\end{array}
\label{eqn:F^2_1}
\end{eqnarray}
where the upper bound for $\theta$ is $\theta= \pi / 2$. Note that the
angular momentum $L$ is unchanged under the inverse map
$\Phi^{-1}$. Similarly, we define the inverse image of $\mathcal F_2^1$  
as  $\mathcal F^2_2 \equiv \Phi^{-1} (\mathcal F^1_2)$, and we have

\begin{widetext}

\begin{figure}
\vspace*{-0.cm} \hspace*{-0.0cm} 
\ig[width=17.5cm,height=7.cm]{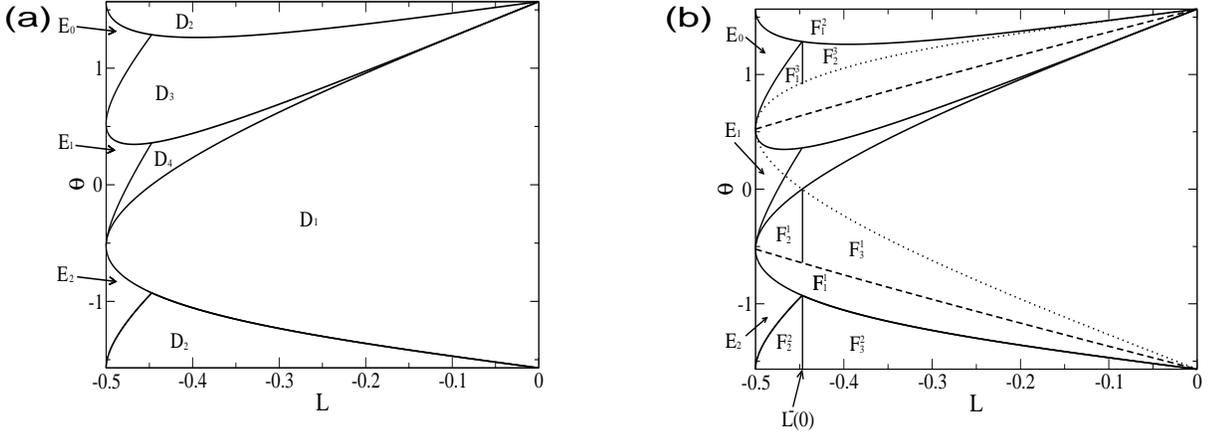}
\vspace*{-0.cm}
 \caption{\label{fig:partition} (a) The first to fourth escape domains 
 $\mathcal D_1$--$\mathcal D_4$ are displayed. The second escape domain
 $\mathcal D_2$ is separated two parts. The remaining part consists of
 three pieces, which we define as $E_0, E_1$ and $E_2$. (b) The regions 
 $\mathcal  F_i^j$ are  displayed. The dotted line indicates the
 boundary of the re-injection domain. } 
\end{figure}

\begin{eqnarray}
\begin{array}{lll}
\mathcal F^2_2 &=&
\{
(L, \theta)~|~\Psi^L(\theta_0(L)) < \theta < \Psi^L (\theta^+(L)),  ~-r < L < L^-(0)
\} \\[.35cm]
&=&
\{
(L, \theta)~|~- \frac {\pi}2 < \theta < 
\theta^+(L) -2 \arcsin \left| L  \right|,  ~-r < L < L^-(0)
\},
\end{array}
\end{eqnarray}
where the lower bound for $\theta$ is $\theta= -\pi /2$.
Finally, let us define the inverse image of $\mathcal F_3^1$ as
$\mathcal F^2_3 \equiv \Phi^{-1} (\mathcal F^1_3)$, and we get
\begin{eqnarray}
\begin{array}{lll}
\mathcal F^2_3 &=&
\{
(L, \theta)~|~
\Psi^L(\theta_0(L)) < \theta < \Psi^L (\theta^-_{\rm in}(L)), 
~L^-(0) < L < 0
\} \\[.35cm]
&=& \{
(L, \theta)~|~- \frac {\pi}2 < \theta < \theta^-(L)
,~ L^-(0) < L < 0
\}
\end{array}
\label{eqn:F23}
\end{eqnarray}
where we have used the relation 
$\theta^-(L) - \theta^-_{\rm in}(L) = -2 \arcsin |L|$.
In the Eq.~(\ref{eqn:F23}), the lower bound for $\theta$ is 
$\theta=  -\pi /2$; the upper is $\theta = \theta^-(L)$, which is
 equivalent to the lower bound of the domain $\mathcal D_1$.
These three sets $\{\mathcal F^2_1, \mathcal F^2_2, \mathcal F^2_3 \}$
 are displayed in Fig.~\ref{fig:partition}(b).  

Next, we derive the third escape domain ${\mathcal D}_3$. It can be proved
 that $\mathcal D_2 \cap \mathcal D_{\rm in}= \phi$, because the 
 relation  
$
\theta^-(L) + \pi - 2 \arcsin \left| L \right| >
\theta_{\rm in}^+ (L)
$
holds.
(See the lower bound of the domain $\mathcal F^2_1$ which is defined by
the second line of the Eq.~(\ref{eqn:F^2_1})). It follows that
$
\mathcal D_3 = \Phi^{-1}(\mathcal D_2) 
= \Phi^{-1} (\mathcal F^2_1) \cup \Phi^{-1} (\mathcal F^2_2) \cup 
  \Phi^{-1} (\mathcal F^2_3)
$
by Eq.~(\ref{eqn:nth-dom}), where the three sets of the right hand side
can be calculated, respectively, as
\begin{eqnarray}
\begin{array}{lll}
\Phi^{-1} (\mathcal F^2_1) 
&=&
\{
(L, \theta)~|~\Psi^L(\theta^-(L) + \pi - 2 \arcsin |L|) 
< \theta < \Psi^L (\frac {\pi}2) 
\} \\[.35cm]
&=&\{
(L, \theta)~|~\theta^-(L) + \pi - 4 \arcsin \left| L \right|
< \theta < \frac {\pi}2 - 2 \arcsin \left| L \right| 
\}, \\[.5cm]
\Phi^{-1} (\mathcal F^2_2) 
& = &
\{
(L, \theta)~|~\Psi^L(-\frac {\pi}2) < \theta < 
\Psi^L (\theta^+(L) -  2 \arcsin \left| L \right| ) 
,~ -r < L < L^-(0)
\} \\[.35cm]
& = &\{
(L, \theta)~|~\frac {\pi}2  - 2 \arcsin \left| L \right| 
< \theta < \theta^+(L) + \pi - 4 \arcsin \left| L  \right|
,~-r < L < L^-(0)
\}, \\[.5cm]
\Phi^{-1} (\mathcal F^2_3) 
&=&
\{
(L, \theta)~|~\Psi^L(- \frac {\pi}2) < \theta < \Psi^L (\theta^-(L)) 
,~ L^-(0) < L < 0
\} \\[.35cm]
& = &\{
(L, \theta)~|~\frac {\pi}2  - 2 \arcsin \left| L \right| 
< \theta < \theta^-(L) + \pi - 2 \arcsin \left| L  \right|
,~L^-(0) < L < 0
\},
\end{array}
\end{eqnarray}
where the relation 
$\Phi^{-1}(\mathcal F^2_1) \subset \mathcal D_{\rm in}$
holds, because of the inequalities
$
\theta_{\rm in}^+(L) > \frac {\pi}2 - 2 \arcsin \left| L  \right|
$
and
$
\theta^-(L) + \pi - 4 \arcsin |L| > \theta^-_{\rm in} (L)
$. 
Therefor, the 4-th escape domain $\mathcal D_4$ is represented by 
$
\mathcal D_4 = \Phi^{-1}(\mathcal D_3 / \mathcal D_{\rm in}) 
= \Phi^{-1} (\mathcal F^3_1) \cup \Phi^{-1} (\mathcal F^3_2),
$
where we define
\begin{eqnarray}
\begin{array}{lll}
\mathcal F^3_1 &=& 
\{ 
(L, \theta)~|~\theta_{\rm in}^+(L) < \theta < 
\theta^+(L) + \pi - 4 \arcsin \left| L \right|, ~-r < L < L^-(0) 
\} \\[.35cm] 
\mathcal F^3_2 &=& 
\{
(L, \theta)~|~\theta_{\rm in}^+(L) < \theta < 
\theta^-(L) + \pi - 2 \arcsin \left| L \right|, ~L^-(0) < L < 0
\}
\end{array}
\end{eqnarray}
These sets $\{\mathcal F^3_1, \mathcal F^3_2 \}$ are displayed in 
Fig.~\ref{fig:partition}(b).  
Thus, the fourth escape domain $\mathcal D_4$ is given by the union of the
 following sets:
\begin{eqnarray}
\begin{array}{lll}
\Phi^{-1} (\mathcal F^3_1) 
&=& \{
(L, \theta)~|~\theta^+(L) 
 < \theta < \theta^+(L) + \pi   - 6 \arcsin \left| L \right| 
,~ -r < L < L^-(0)
\} \label{eqn:F^4_2} \\[.35cm]
\Phi^{-1} (\mathcal F^3_2) 
&=& \{
(L, \theta)~|~ \theta^+(L)
 < \theta < \theta^-(L) + \pi   - 4 \arcsin \left| L \right| 
,~ L^-(0) < L < 0
\},
\end{array}
\end{eqnarray}
where we have used the relation 
$\theta^+(L) - \theta^+_{\rm in}(L) = -2 \arcsin |L|$.
The lower bounds for $\theta$ of these two sets are
 $\theta=\theta^+(L)$; and the upper bounds of the set 
$\Phi^{-1} (\mathcal F^3_2)$ is equivalent to the lower bound of
 $\mathcal D_3$. 

\begin{figure}
 \vspace*{-.cm} \hspace*{-0.0cm}
\ig[width=17.5cm,height=6.5cm]{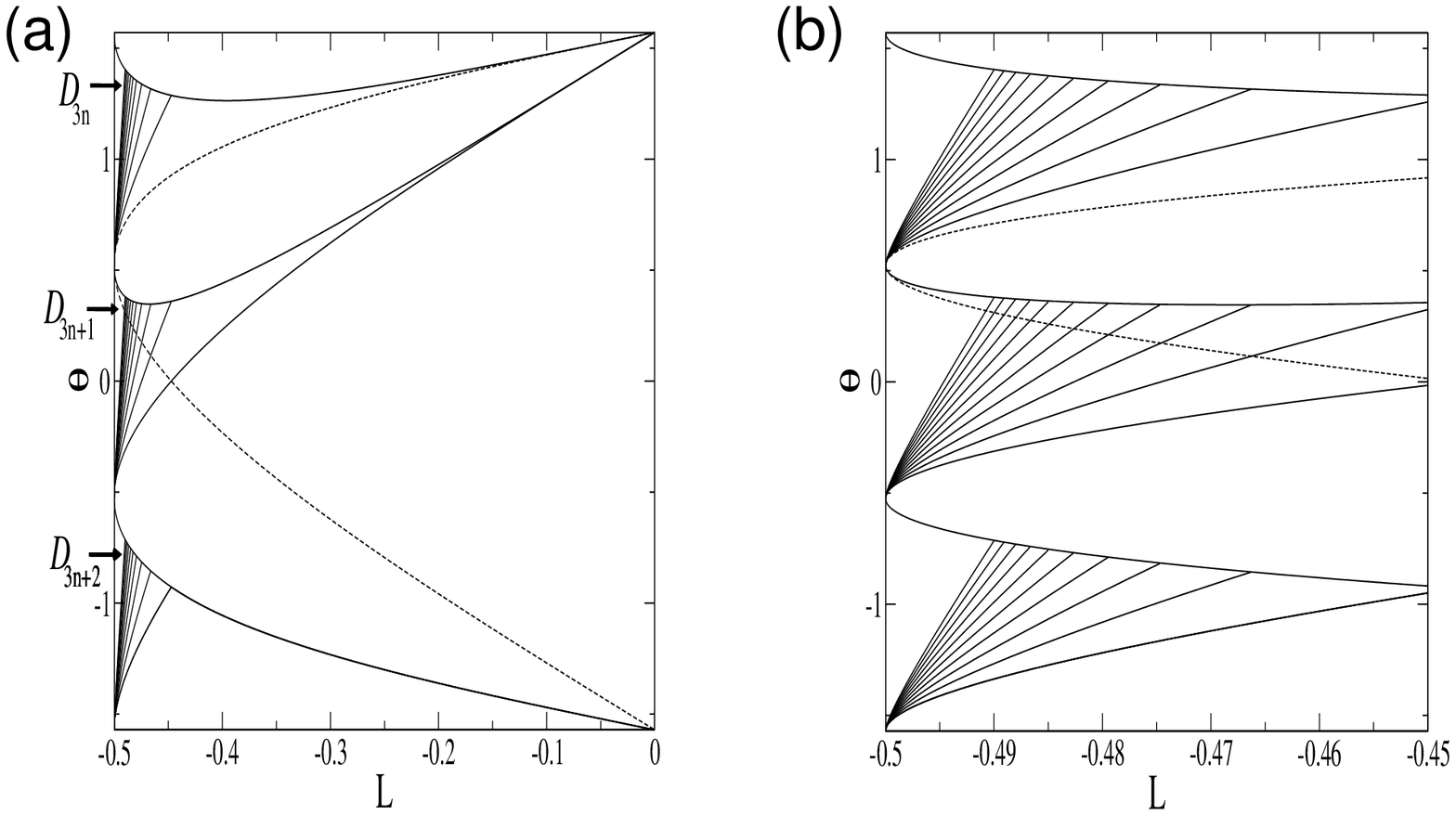}
 \vspace*{.3cm}
 \caption{\label{fig:partition2}(a) The infinite partition
 constructed in terms of the escape time. The solid lines represents
 boundaries between regions of different escape times. The right hand
 side of the broken line is the injection domain.  (b) A magnification
 of Fig.~\ref{fig:partition2}(a) in a neighborhood  of the outermost 
 tori ($L=-0.5$). The boundaries for the domains with the escape times
 longer than 32 are omitted. 
 } 
\end{figure}

\end{widetext}

From the above results, the domain ${\mathcal D^-}$ are covered by the
four sets $\mathcal D_1, \mathcal D_2, \mathcal D_3$, and $\mathcal D_4$
except for the three regions $E_0,E_1$, and $ E_2$ as illustrated in
Fig.~\ref{fig:partition}(a)(b). 
The boundaries of the $n$-th escape domain ($n \geq 5$) can be obtained
by recursively calculating the inverse mapping $\Phi^{-1}$ of the upper
bound of the set $\Phi^{-1} (\mathcal F_1^3)$, which is given by
Eq.~(\ref{eqn:F^4_2}) as $\theta=\theta^+(L) + \pi - 6 \arcsin |L|$.
Thus, let us define the boundary between the domains $\mathcal D_{3n+1}$
and $\mathcal D_{3(n+1)+1}$ as $\theta_{3n+1}(L)$ ($n \geq 1$), we have   
\begin{eqnarray}
\theta_{\rm 3n+1} (L) 
= \theta^+ (L) + n \pi - 6n \arcsin \left| L \right|  
\label{eqn:theta-3n+1}
\end{eqnarray}
Similarly, defining the boundaries between $\mathcal D_{3n+2}$ and 
$\mathcal D_{3(n+1)+2}$ as $\theta_{3n+2}(L)$, and between 
$\mathcal D_{3n}$ and $\mathcal D_{3(n+1)}$ as  $\theta_{3n}(L)$~,
we have 
\begin{eqnarray}
\begin{array}{lll}
\theta_{\rm 3n+2} (L) 
&=& \theta^+ (L) + n \pi - 2(3n+1) \arcsin \left| L \right|, \\[.35cm]
\theta_{\rm 3n} (L) 
&=& \theta^+ (L) + n \pi - 2(3n-1) \arcsin \left| L \right| 
\end{array}
\end{eqnarray}
In Figs.~\ref{fig:partition2}(a) and (b),  we
depict these boundaries up to $n=31$. 

\section{The escape time distribution}
\begin{figure}
\vspace*{-.cm}
\ig[width=8.5cm,height=6.5cm]{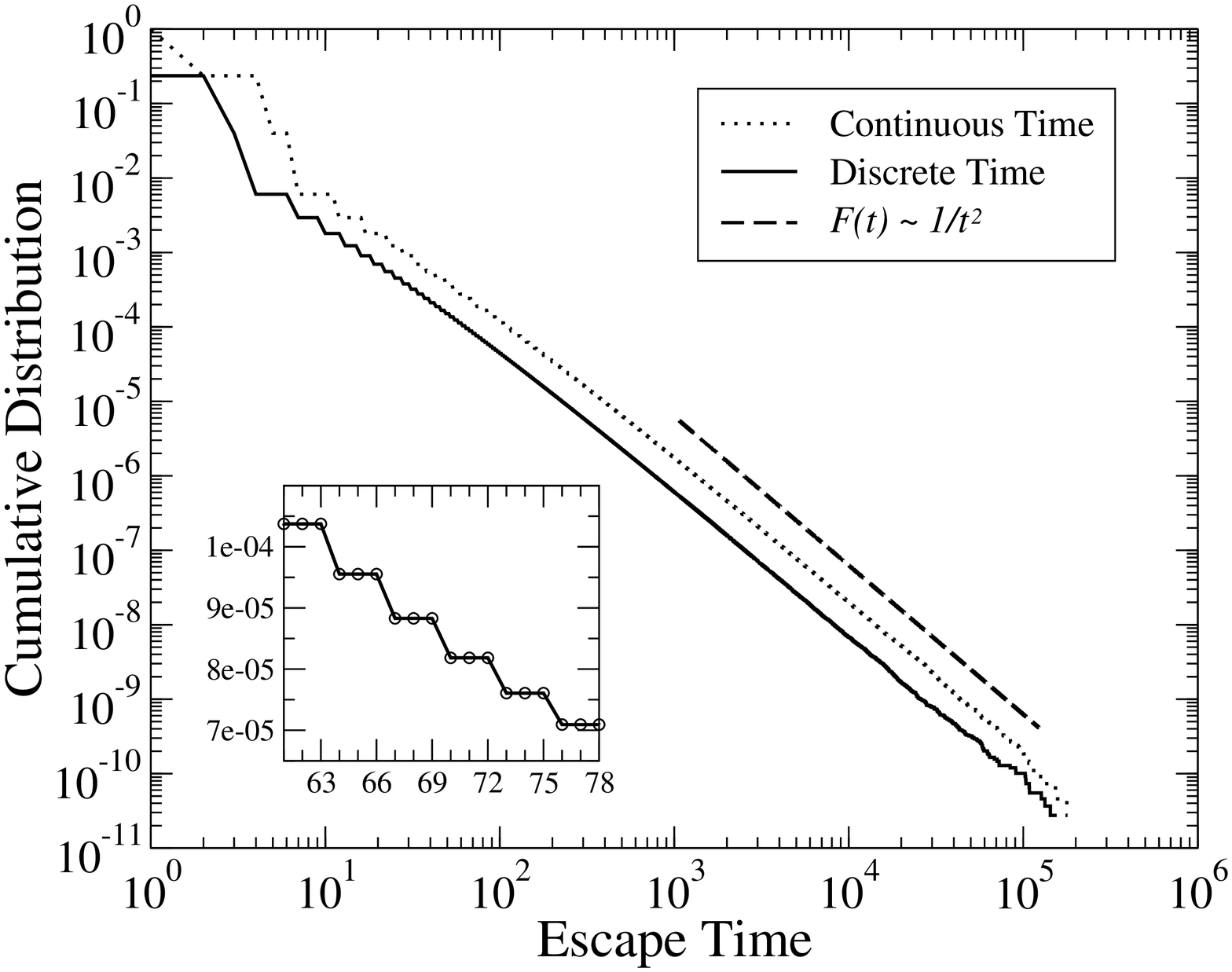}
\vspace*{-.cm}
 \caption{\label{fig:esc-time} The cumulative distributions of the
 escape times for discrete time (the solid line) and for continuous time  
 (the dotted line) in log-log form. The dashed line is the analytical
 result $F(t) \sim 1/t^2$. The inset is a magnification of the discrete
 time case, which shows a clear stepwise structure.
}
\end{figure}

Finally, we derive a scaling property of the escape time
distribution $f_{\rm esc}(n)$ approximately. The escape time is defined 
by the number of collisions with the arc of the semicircle just after an  
orbit enters the hat region until it escapes from there.

Since the Poincar\'e map is area-preserving, which is the universal
property of the Poincar\'e map of the Hamiltonian systems, the 
physically natural invariant measure of the Poincar\'e map is the
Lebesgue measure. Thus the probability that the escape time equals $n$
is given by the Lebesgue measure $S_n^{\rm in}$ of the region 
$\mathcal D_{n} \cap \mathcal D_{\rm in}$, that is 
$f_{\rm esc}(n) = S_n^{\rm in}$.  Note that 
$S_{3n}^{\rm in} = S_{3n-1}^{\rm in} = 0 ~(n=2,3,\cdots)$ and thus only  
$S_{3n+1}^{\rm in}~(n=1,2,\cdots)$ has finite values (See
Fig.~\ref{fig:partition2}(a)(b).).   

\begin{figure}
\vspace*{-.3cm}
\ig[width=8.5cm,height=7.0cm]{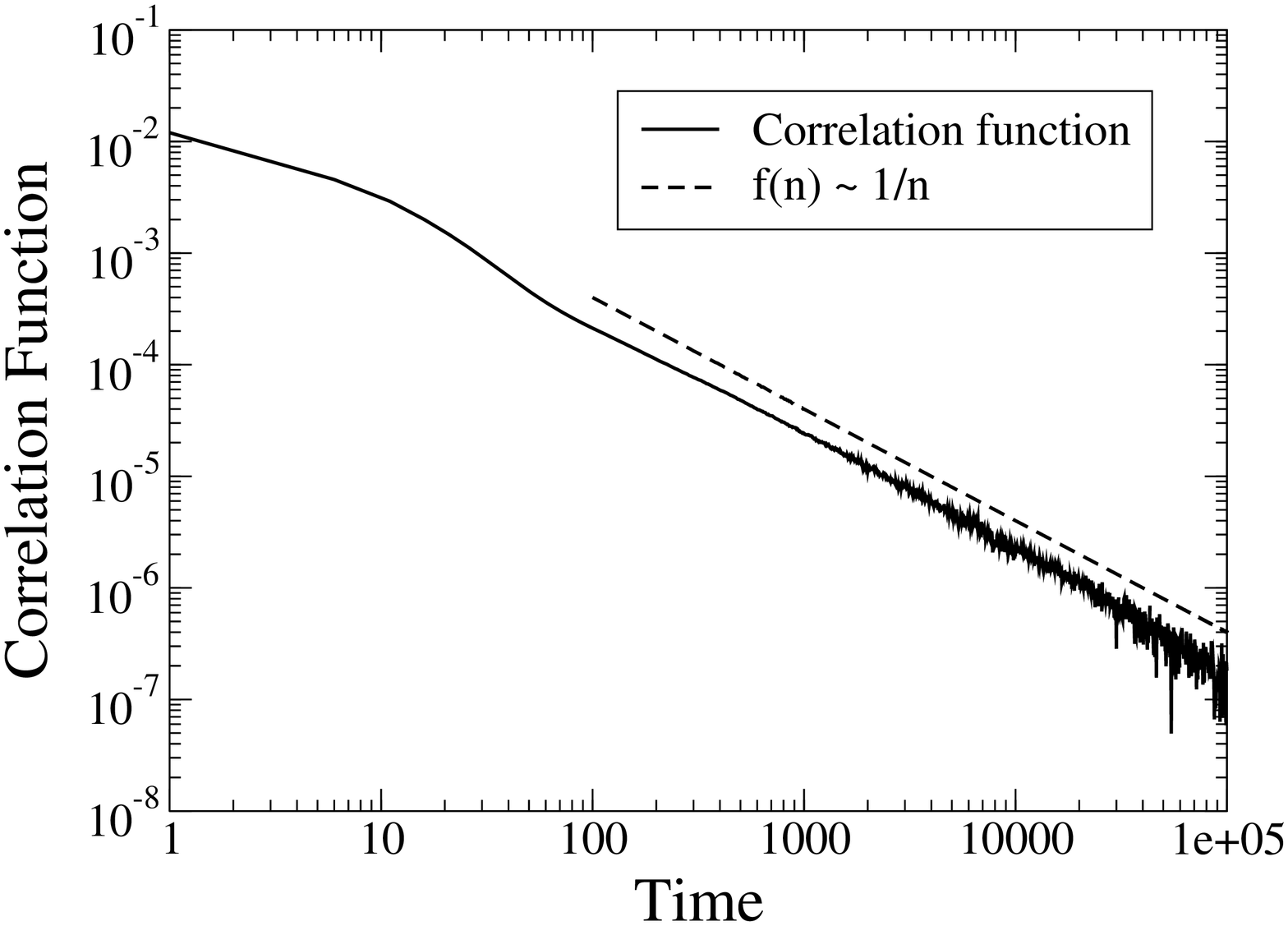}
\vspace*{-.2cm}
 \caption{\label{fig:corr} Auto correlation function of the absolute
 value of the angular momentum $|L| - \langle |L| \rangle$ (the solid
 line) in log-log form, where $\langle \cdot \rangle$ means the ensemble
 average in terms of the Lebesgue measure in the chaotic domain. The
 broken line represents a function $f(n) \sim 1/n$, which is a guide to
 see.}    
\end{figure}

Let us derive the cross points of the lines  
$
\theta_{3n+1}(L) = 
\theta^+(L) + n \pi - 6 n \arcsin \left| L \right| 
$
[Eq.~(\ref{eqn:theta-3n+1})] and $\theta (L) = 0$. Using the Taylor
expansions 
$\arcsin x \approx \pm \frac {\pi}6 
 + \frac 2{\sqrt 3} \left(x \mp \frac 12\right)$
~~(as $x \approx \pm \frac 12$),
we have
\begin{eqnarray}
\begin{array}{lll}
\theta_{3n+1}(L)  &\approx&  \\[.35cm]
&& \hspace*{-1.3cm} 
- \f {\pi}6 +
\f 2{\s 3} (3n+1)(2L+1) +
 \s 2 (2L+1)^{\frac 12},
\end{array}
\end{eqnarray}
as $n \to \infty$. Setting this to $0$, and solving in terms of $L$, we
have $2L+1 \sim  \frac 1{3n+1} \sim \frac 1n$. Thus we find the width
of the $(3n+1)$-th escape domain is proportional to $1/n^2$. It follows
that the area of the $n$-th escape domain $S_n$ behaves as 
\begin{eqnarray}
S_n \sim \f 1{n^2},
\label{eqn:sc}
\end{eqnarray}
as $n \to \infty$. Note that $S_{3n+1} = S_{3n} = S_{3n-1}$ for
$n=2,3,\cdots$, because the domains $\mathcal D_{3n}$ and 
$\mathcal D_{3n-1}$ has no intersection with the domain 
$\mathcal D_{\rm in}$. The Eq.~(\ref{eqn:sc}) means the fact that 
the partition constructed in the previous section is infinite.   
Finally, we get 
\begin{eqnarray}
S_{3n+1}^{\rm in} = S_{3n+1} - S_{3n+4} \sim \f 1{n^3},
\end{eqnarray}
as $n \to \infty$. And, as mentioned above, the equations 
$S_{3n}^{\rm in} = S_{3n-1}^{\rm in} = 0$ hold.
This power law perfectly agrees with the numerical results shown in
Fig.~\ref{fig:esc-time}, where the cumulative distribution 
\begin{eqnarray}
F_{\rm esc}(n) := \sum_{j=n+1}^{\infty} f_{\rm esc}(j)
\end{eqnarray}
is plotted by the solid line. Note that this numerical result have
already been reported by Altmann et. al.\cite{altmann}.  In the inset, a
magnification is displayed, which shows a clear stepwise structure with
decreases exactly at $n = 3k+1~(k=1,2,\cdots)$. This implies that
$f_{\rm esc}(3k) = f_{\rm esc}(3k+2) = 0$, and that the only 
$f_{\rm esc}(3k+1)$ have finite values $(k \geq 2)$. Thus, these results
also agree with the analytical results.

\section{Concluding remarks}

In conclusion, we have derived the escape time distribution by
constructing the infinite partition in terms of the escape times. 
Note that, however, the escape 'time' in this paper is the number of
collision until the particle escapes. Thus the escape time of the
continuous time flow might be slightly different from ours. But the
scaling exponents should be the same, because the flight time of the
chaotic orbits between  collisions in the hat region is
non-vanishing. This is confirmed numerically and the result is displayed 
in Fig.~\ref{fig:esc-time} which shows the agreement of the scaling 
exponents of these two distributions. 

There are several points that should be verified in future studies.
First, the correlation functions of this system exhibits power law
behavior (Fig.~\ref{fig:corr}, see also \cite{bunimovich}), and it is
expected that there are relations between the scaling exponent of the
escape time distributions and the correlation functions. 
Second, it is important to elucidate whether the results in this paper 
is general or not for other parameter values $R \ne 2r$. Third, it
is also important to consider which property of the mushroom billiard is
the universal features of the mixed type Hamiltonian systems.

\begin{acknowledgments}
 I would like to thanks Prof.~Y.~Aizawa and Prof.~A.~Shudo for
 valuable discussions and comments. This work is supported in part by
 Waseda University Grant for Special Research Projects (The Individual
 Research No.~2005B-243) from Waseda University.    
\end{acknowledgments}

\bibliography{mushroom}

\end{document}